\documentclass[preprint,5p,times,twocolumn]{elsarticle}
\usepackage{verbatim}
\usepackage{listings}
\usepackage{multirow}
\usepackage{graphicx} 
\usepackage{lineno}
\usepackage{array}
\usepackage{hyperref} 
\usepackage{graphicx}
\usepackage{amsmath,amssymb,amsthm}
\usepackage{rotating}
\usepackage{subfigure} %
\def\tect{$^{130}$Te }
\def\tectn{$^{130}$Te}

\def\pbdd{$^{210}$Pb }
\def\pbddn{$^{210}$Pb}
\def\podd{$^{210}$Po }
\def\poddn{$^{210}$Po}
\def\ptc{$^{190}$Pt }
\def\udt{$^{238}$U }
\def\udtn{$^{238}$U}
\def\thdt{$^{232}$Th }
\def\thdtn{$^{232}$Th}
\def\tld{$^{208}$Tl }

\def\kq{$^{40}$K }

\def\coss{$^{60}$Co }

\def\BBz{$\beta\beta0\nu$}
\def\BBzn{$\beta\beta0\nu$~}

\def\ca{$\sim$}

\def\pom{$\pm$ }

\def\teod{TeO$_2$~}
\def\teodn{TeO$_2$}
\def\be{\begin{equation}}
\def\ee{\end{equation}}

\def\ciccio{5$\times$5$\times$5 cm$^3$ }

\def\0nbb{$0\nu\beta\beta$}

\def\Cplusplus{C\raise.5ex\hbox{\small ++}}
 
\begin{document}
\newenvironment{shell}{\vspace{5mm}\newline}{\vspace{5mm}\newline}
\newenvironment{code}{\vspace{5mm}\footnotesize\verbatim}{\endverbatim\normalsize\vspace{5mm}} 
\title{Validation of techniques to mitigate copper surface contamination in CUORE}

\author[INFNMilano]{F.~Alessandria}
\author[MilanoPoli]{R.~Ardito}
\author[USC,LNGS]{D.~R.~Artusa}
\author[USC]{F.~T.~Avignone~III}
\author[INFNLegnaro]{O.~Azzolini}
\author[LNGS]{M.~Balata}
\author[BerkeleyPhys,LBNLNucSci,LNGS]{T.~I.~Banks}
\author[INFNBologna]{G.~Bari}
\author[LBNLMatSci]{J.~Beeman}
\author[Roma,INFNRoma]{F.~Bellini}
% ...

\address[INFNMilano]{INFN - Sezione di Milano, Milano I-20133 - Italy}
\address[MilanoPoli]{Dipartimento di Ingegneria Strutturale, Politecnico di Milano, Milano I-20133 - Italy}
\address[USC]{Department of Physics and Astronomy, University of South Carolina, Columbia, SC 29208 - USA}
\address[LNGS]{INFN - Laboratori Nazionali del Gran Sasso, Assergi (L'Aquila) I-67010 - Italy}
\address[INFNLegnaro]{INFN - Laboratori Nazionali di Legnaro, Legnaro (Padova) I-35020 - Italy}
\address[BerkeleyPhys]{Department of Physics, University of California, Berkeley, CA 94720 - USA}
\address[LBNLNucSci]{Nuclear Science Division, Lawrence Berkeley National Laboratory, Berkeley, CA 94720 - USA}
\address[INFNBologna]{INFN - Sezione di Bologna, Bologna I-40127 - Italy}
\address[LBNLMatSci]{Materials Science Division, Lawrence Berkeley National Laboratory, Berkeley, CA 94720 - USA}
\address[Roma]{Dipartimento di Fisica, Sapienza Universit\`a di Roma, Roma I-00185 - Italy }
\address[INFNRoma]{INFN - Sezione di Roma, Roma I-00185 - Italy }
% ...

\begin{abstract}
In this article we describe the background challenges for the CUORE experiment posed by surface contamination of inert detector materials such as copper, and present three techniques explored to mitigate these backgrounds.  Using data from a dedicated test apparatus constructed to validate and compare these techniques we demonstrate that copper surface contamination levels better than 10$^{-7}$ - 10$^{-8}$~Bq/cm$^2$ are achieved for \udt and \thdtn.
If these levels are reproduced in the final CUORE apparatus the projected 90\%~C.L. upper limit on the number of background counts in the region of interest is 0.02-0.03~counts/keV/kg/y depending on the adopted mitigation technique.
\end{abstract}

\maketitle


\section{Introduction}
\label{sec:intro}
Neutrinoless Double Beta Decay (\BBz) is an area of vigorous experimental activity with potential for profound impact on modern questions in fundamental physics~\cite{02elliot,08avignone,10barabash}. Observation of this decay would immediately imply lepton number violation and would establish the neutrino as a Majorana fermion.  If neutrinos are indeed Majorana fermions, a measurement of the \BBz\ rate would probe the absolute neutrino mass scale and possibly reveal the neutrino mass hierarchy. 
 
Sensitivity to the non-degenerate inverted hierarchy of neutrino masses is a standard benchmark for next generation \BBzn\ searches which demands that very low background levels -- of the order of few counts per ton per year in the region of interest -- be achieved.  All aspects of the experiment, for example selection of materials, machining and handling of components, and assembly procedures must be scrutinized for background control.  Validation of effective control measures and quantifying the residual background is often nearly as challenging as the underlying experiment.   This paper focuses on aspects of the background control and validation activities for the CUORE experiment~\cite{03alessandrello,03arnaboldi,04arnaboldi, 09pavan}. Specifically, we present a study of three techniques explored to mitigate background from residual surface radioactivity on structural materials in the detector, particularly copper.

\section{Overview of the CUORE detector}
\label{sec:CUORE_bkg}

The CUORE experiment, currently under construction underground at Laboratori Nazionali del Gran Sasso (LNGS), will search for \BBzn of \tectn. The signature of this decay is a peak in the energy spectrum centered at the Q-value of the transition, at about 2528~keV~\cite{09redshaw,09scielzo,11rahaman}. The experimental goals include a background level of  $\leq$10$^{-2}$ counts/keV/kg/y in an energy window of \ca 100~keV around the Q-value, denoted the region of interest (ROI), and a high-precision measurement of the spectrum in that region. 

\begin{figure}[tp] 
\centering
\includegraphics[trim=0in 0in 0in 0in, clip=true, height=0.3\textwidth]{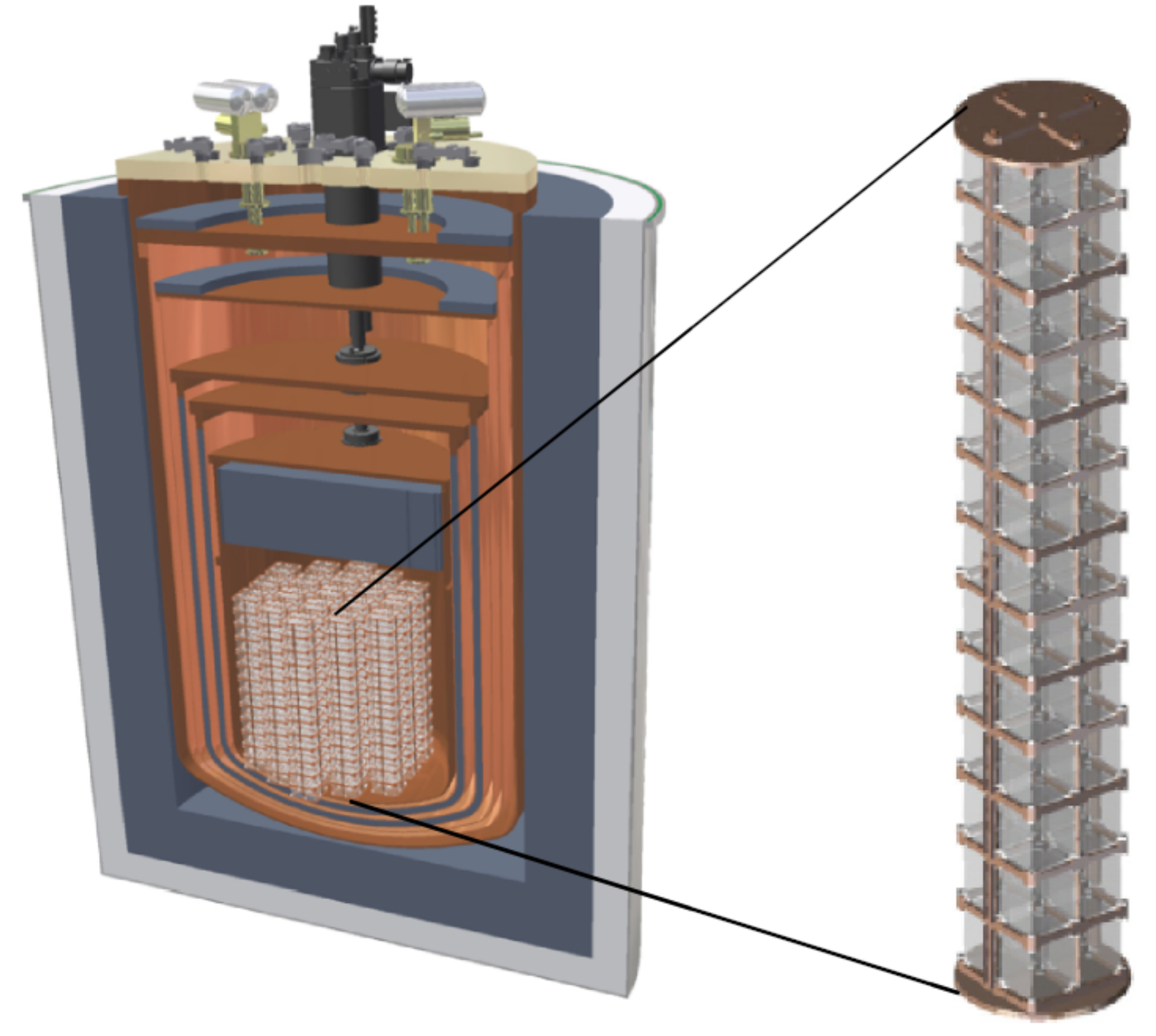}%
\hspace{0.01\textwidth}%
\caption[Set-up of the CUORE experiment]{Set-up of the CUORE experiment: the 988
bolometers arranged in a 19 towers array, hanging in vacuum inside nested copper
cylindrical vessels and provided with lead shields. On the right a detail of one CUORE tower.}
\label{fig:apparatus}
\end{figure}

The apparatus, shown in Figure~\ref{fig:apparatus} will consist of a close-packed array of 988, $5\times5\times5$~cm$^3$ cubic \teod crystals, amounting to 206~kg of \tectn. These will be cooled inside a cryostat to around 10~mK. At this temperature the crystals function as highly sensitive calorimeters, converting the energy deposited in their volume to a measurable temperature change. The bolometers will be arranged in a compact cylindrical matrix of 19 towers, each tower will contain 13 planes of four crystals. A copper skeleton will provide the mechanical structure to hold the crystals in each tower. The array will hang in vacuum inside a copper cylindrical vessel closed on the top and bottom with copper plates. The copper skeleton, denoted collectively as the \emph{copper holder}, will not touch the crystals directly, instead PTFE standoffs will secure the crystals.  Components made of material other than copper or \teod make up a small fraction of the detector. These components, denoted collectively as \emph{small parts}, include the thermistors used to read out the bolometric signal, the silicon heaters, used to check for gain variations, the glue used to attach the thermistors to the crystals, the PTFE standoffs, and the readout wires.  A complete description of the CUORE detector can be found in~\cite{04arnaboldi}.

\begin{figure*}[tp] 
\centering
\includegraphics[trim=0in 0in 0in 0in, clip=true, width=0.45\textwidth]{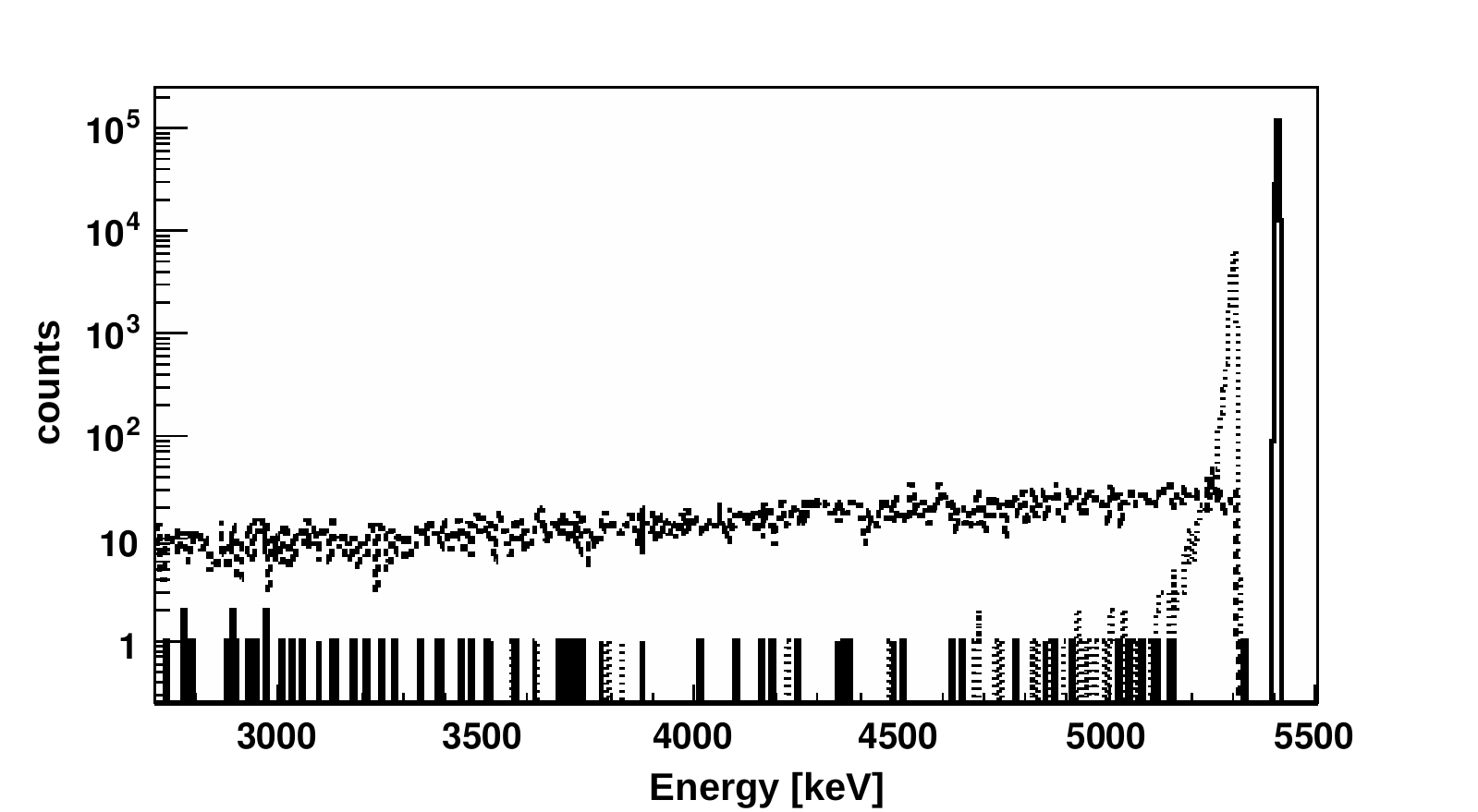}%
\hspace{0.01\textwidth}%
\includegraphics[trim=0in 0in 0in 0in, clip=true, width=0.45\textwidth]{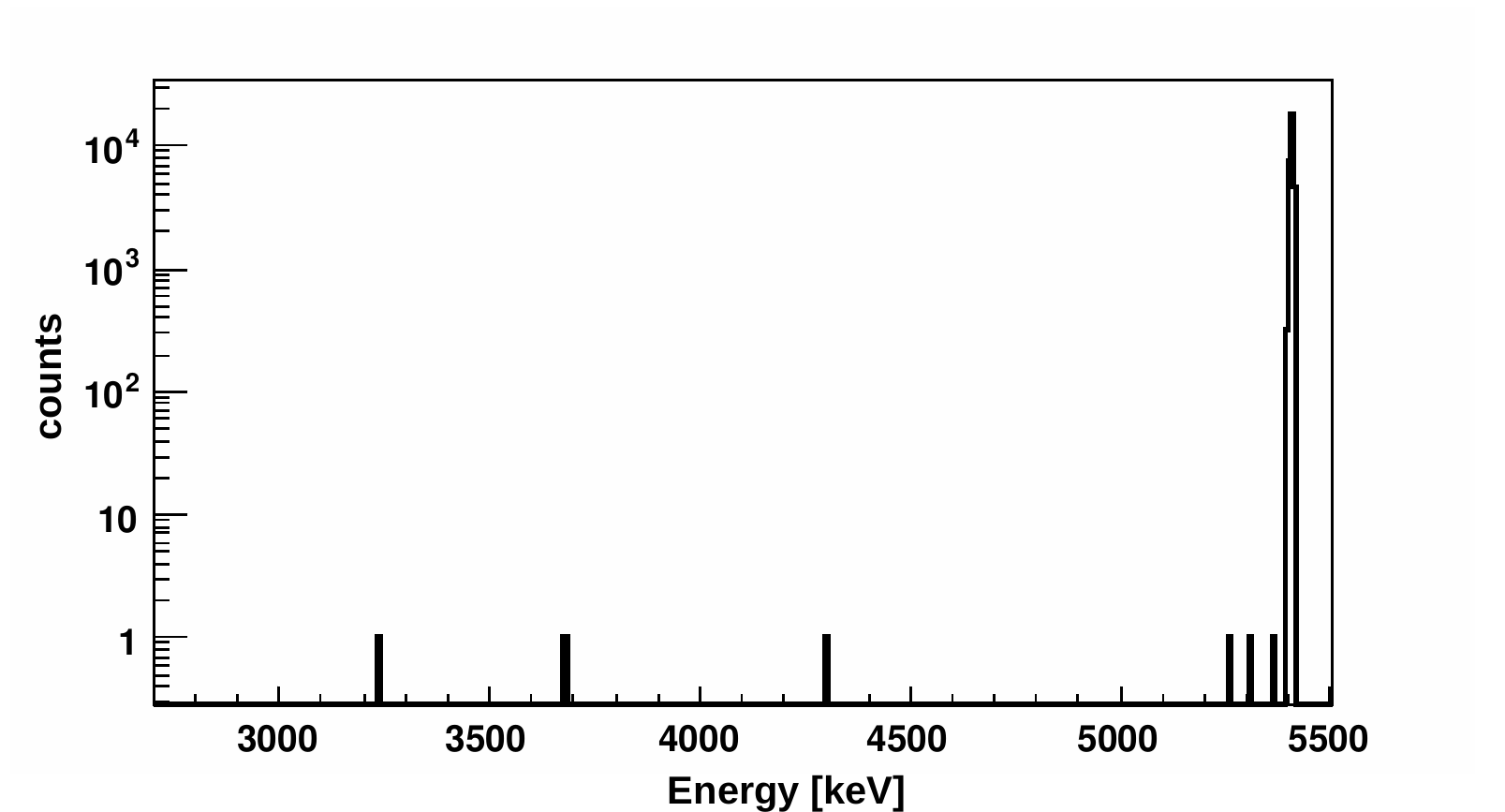}%
\hspace{0.01\textwidth}%
\caption[Monte Carlo simulation spectra for \podd $\alpha$ decay in the crystal bulk/surface and on the copper surface]{Left plot: multiplicity=1 spectra from Monte Carlo simulations of \podd $\alpha$ decay in the crystal bulk (solid-line histogram), spread on a very shallow depth (dotted-line histogram) and on a deeper depth (dashed-line histogram) of copper surface. Right plot: Monte Carlo simulated spectrum obtained summing up the energy released by two events in coincidence for a \podd surface decay on the crystals.}
\label{fig:MC}
\end{figure*}


\section{Bolometer background from  surface and bulk radioactivity}
\label{sec:MC}

In this Section we discuss aspects of bolometer behaviour that influence their susceptability to background from decays of surface radio-impurities and describe some discriminators to distinguish between surface and bulk contamination.  In the context of this paper, bulk contamination represents unwanted impurities distributed uniformly throughout the volume of a material. This contamination arises from impurities present in the raw material, or introduced during manufacturing.  On the other hand surface contamination refers to impurities from the environment that adhere to, or become embedded in the exposed surfaces of the material or to bulk impurities diffused into the surface.  Since some impurities, for example long-lived decay fragments of \udt or \thdtn, can become implanted inside the surface of the material the contamination is not strictly a surface distribution but rather may be distributed in a thin layer near the surface. 

Although bolometers make excellent calorimeters~\cite{00alessandrello,05arnaboldi_bolo}, achieving FWHM energy resolution as high as 3.9 \pom 0.7~keV at 2614.5~keV~\cite{00alessandrello}, the phonon signal at the heart of the method shows very little dependence on either the identity of the particle responsible for the underlying energy deposition or on the position of the event in the crystal. Electrons, gammas, $\alpha$-particles and nuclear recoils depositing the same amount of energy at any point in the bolometer produce virtually identical pulses~\cite{97alessandrello, 10belliniQF}. The lack of particle identification and the absence of a surface dead-layer on the crystals means that $\alpha$'s as well as $\beta$'s and $\gamma$'s form a background if they deposit an amount of energy in the bolometer equivalent to the \BBzn transition energy.

Only a few decays in the chains of \udt and \thdt emit $\beta$- and $\gamma$-particles with sufficient energy to mimic a \BBzn candidate, the most dangerous being the 2614.5~keV $\gamma$-ray from \tld --- a daughter of \thdtn. 
On the other hand all $\alpha$ decays in the \udt and \thdt chains involve transition energies much higher than the \tect Q-value; if the entire transition energy is absorbed in one bolometer these do not contribute to the \BBzn background. However, this is not true if the $\alpha$ decay occurs on the surface of the detector or of the facing inert materials. In this case the transition energy is shared among more than one element and if the correct amount of energy is deposited in any crystal on the trajectory of the decay fragments, the pulse from that crystal could mimic a \BBzn candidate. Thus degraded $\alpha$'s from surface impurities form a potentially serious background for bolometers. Studies performed on data acquired with bolometric experiments preceeding CUORE -- like MiDBD~\cite{03arnaboldi_array20} and Cuoricino~\cite{05arnaboldi} -- indicated surface contaminations as the most pernicious source of background~\cite{09bucci, 08arnaboldi}.
  
A very useful parameter to identify and veto decays from surface contamination is {\it event multiplicity}. This is defined as the number of crystals firing in coincidence, typically within 100~ms of each other (typical rise and decay times are of the order of tents and hundreds of ms respectively, and the expected average rate is around a few mHz). Particles emitted from decay of surface contamination can hit more than one bolometer, whereas a true \BBzn candidate is expected to be contained within a single crystal with an efficiency of \ca83\%~\cite{10andreotti}.
As an example, we show in Figure~\ref{fig:MC} and discuss below, results from a Monte Carlo (MC) simulation which illustrate the multiplicity characteristics of $\alpha$ decay of $^{210}$Po, a daughter of $^{210}$Pb.

First we consider if the \podd decays in the bulk or in the surface of \teod crystals.
In the bulk, the energy of both the $\alpha$ and recoiling nucleus are absorbed fully in the crystal. Thus a gaussian peak centered at the isotope transition energy in the multiplicity=1 (M1) spectrum is found, see the solid-line histogram in the left plot of Figure~\ref{fig:MC}. 

On the other hand, in the surface region of a crystal, $\alpha$'s - or/and the recoiling nuclei - are not fully stopped in one crystal and sometimes enter a second bolometer causing both to fire in coincidence. The case where two bolometers fire in coincidence corresponds to multiplicity=2 (M2) events. The summed energy of the two events is denoted M2Sum. The M2Sum spectrum has a gaussian peak centered at the full isotope transition energy since the $\alpha$ and recoiling nucleus are fully absorbed in the two crystals. See right plot of Figure~\ref{fig:MC}. For more details see Ref.~\cite{11CCVR}.

Next we consider the case of contamination in the surface region of inert materials such as copper. Degraded $\alpha$'s may escape the copper and hit a bolometer. Only the energy deposited in the bolometer can be observed, the energy deposited in the inert material is lost. For very shallow surface contamination a peak at the $\alpha$ energy --- the transition energy minus the nuclear recoil energy --- is visible in M1 spectrum, see the dotted-line histogram in the left plot of Figure~\ref{fig:MC}.   However, when the contamination is deeper, the range of energies possible for the degraded $\alpha$ becomes broader and an almost flat continuum, extending from the full $\alpha$ energy down to the lowest energies, is found. For example, see the dashed-line histogram in the left plot of Figure~\ref{fig:MC}. 
  
The energy interval from 2.7-3.9~MeV is an ideal region to study the degraded $\alpha$ background as it is above the highest $\gamma$-line in the \udt and \thdt chains,  2614.5~keV, but below the lowest $\alpha$-line at about 3947~MeV from \thdtn. The only peak visible in \teod experimental spectra in this region is at 3270~keV,   this line comes from $^{190}$Pt bulk contamination of \teod crystals due to inclusion of platinum atoms from the crucible during crystal growth~\cite{10TeO2}.  In the subsequent Sections we refer to the energy interval from 2.7~MeV to 3.9~MeV, excluding the 200~keV window centered on the $^{190}$Pt peak, as the {\it degraded $\alpha$ window}.

The Cuoricino experiment~\cite{05arnaboldi, 08arnaboldi}, a precursor to CUORE, found that the observed background in the ROI for the \BBzn of \tectn, 0.18 $\pm$ 0.01 counts/keV/kg/y, was consistent with the expectation from three classes of sources: (i) multi-Compton events caused by 2614.5~keV $\gamma$-radiation from \tld supported by \thdt contamination of the cryostat or its shields; (ii) radiation from surface contamination of the \teod crystals with \thdt or \udtn~\footnote[1]{Throughout this paper we focus on decays in the chains of the long-lived naturally occurring isotopes: {\thdtn}, {\udtn}. \poddn, a daughter of \pbddn, is also studied as an independent source as it is often found out of secular equilibrium. The source could be both a \pbdd and a \podd contamination. Unfortunately, due to the high energy threshold (above the 46~keV \pbdd $\gamma$ line), we have no chance to distinguish among them. In the text we will therefore refer to \podd since the studied signature is its $\alpha$ peak.}; and (iii)  radiation from surface contamination of inert materials surrounding the crystals, most likely copper. In the case (ii) and (iii) the radiation is primarily degraded $\alpha$'s with a small contribution from $\beta$+$\gamma$ events. Using signatures from outside the ROI to fix the normalization of each source the extrapolated count rate in the ROI was found equivalent to: (i) $30 \pm 10$\%, (ii) $10 \pm 5$\%  and (iii) $50 \pm 20$\%  of the observed background rate. Although this finding does not exclude other sub-dominant backgrounds, the study indicates that \ca 30\% of the background originated from the cryogenic apparatus and \ca 70\% from detector surface contamination, particularly degraded $\alpha$'s.  

To meet the ROI-background goal of CUORE both these sources must be addressed. Materials for the CUORE cryostat and the internal and external shields with sufficient radiopurity to ensure a ROI-background contribution far below the target level of 10$^{-2}$~ counts/keV/kg/y have already been identified~\cite{04arnaboldi}.
For the \teod crystals and the copper holder the acceptable \thdt and \udt contamination levels are as low as 10(\thdtn)-100(\udtn) $\mu$Bq/kg for the bulk and 1-10 nBq/cm$^2$ for the surface.  While copper bulk contamination can be adequately validated with HPGe spectroscopy and Neutron Activation Analysis (NAA) (90\%~C.L. upper limits of 2  $\mu$Bq/kg for \thdt~\cite{11Max} and of 65 $\mu$Bq/kg for \udt have been obtained for CUORE copper samples with NAA and HPGe measurements respectively), these techniques do not have sufficient sensitivity to validate the radiopurity of the copper surface, the \teod surface, and the \teod bulk. 

To meet this validation challenge a series of test arrays containing  a few bolometers arranged in the style of one or a few floors of CUORE were operated in dedicated runs at LNGS~\cite{09pavan}.  The results of \teod crystal surface validation runs were reported in~\cite{11CCVR}; 90\%~C.L. upper limits on \udt and \thdt surface concentration corresponding to 3.8~nBq/cm$^2$ and 2.0~nBq/cm$^2$ respectively were demonstrated. 
The role of copper and the results and interpretation of test runs dedicated to copper surface validation are discussed in the remainder of this paper. We focus exclusively on copper since next to \teod this makes up the largest material mass and surface in the detector. 
Due to the small mass and surface area of the detector small parts, the request on their radiopurity is less stringent than for copper. The sensitivity of standard spectroscopic techniques, like HPGe (for bulk contamination), and Si barrier detector (for surface contamination), is therefore enough to exclude a significant contribution of the detector small parts to the  ROI-background.
 
\section{Copper contamination and treatment for CUORE}\label{sec:cuCUORE}
The mechanical and cryogenic properties of copper make it an ideal material for CUORE.  NOSV copper~\cite{CuNOSV}, produced by the Norddeutsche Affinerie~\cite{NHA}, was selected for the experiment because of its high residual-resistivity ratio (RRR) (certified to be higher than 400) --- a constraint from our cryogenic application --- and because samples were found to be extremely radio-pure. 

Cosmogenic activation of the copper, especially \coss production, can  be controlled to an acceptable level by storing it underground at LNGS except for the time needed for machining and cleaning.  We estimate the time above ground of about 4 months from cast to final storage.  With this exposure time, using the production cross sections in~\cite{cosmo}, we expect the cosmogenic activity of the copper holder will be less than 50 $\mu$Bq/kg when CUORE starts taking data, ensuring a negligible contribution to the background in the \BBz-ROI.

Although the copper bulk-contamination satisfies the requirements of CUORE the surface contamination of commercially available copper needs to be reduced.
Unfortunately standard surface cleaning procedures, generally based on pickling and etching, appear to be unable to reach the desired radiopurity.
For Cuoricino the surfaces of all the copper components facing the bolometers were treated at Laboratori Nazionali di Legnaro (LNL) with a procedure similar to one used for resonant cavity production~\cite{ResonantCavity1,ResonantCavity2}. It involved a sequence of tumbling, to reduce the surface roughness; chemical etching, to reduce chemical contamination; and passivation, to decrease the possibility of re-contamination. Although this procedure proved more effective than standard methods~\cite{03arnaboldi_array20,09bucci},  the rate in the ROI due to contamination of copper surfaces in Cuoricino was constrained to be 50 $\pm$ 20\% of the observed ROI-background rate~\footnote[2]{Important environmental muon and neutron contribuions were ruled-out on the basis of simulations and measurements~\cite{10bellini,09bucci,10andreotti_mu}.}.

To improve on this for CUORE we strive to both minimize the amount of copper and other inert materials facing the bolometers and to identify better surface treatment to mitigate the background from the remaining surfaces.  Three surface treatments techniques were chosen to test: (i) wrapping of surfaces with polyethylene, (ii) simple surface cleaning with ultra-clean acids, and (iii) a modification of the LNL procedure. A dedicated bolometric test, called the Three Towers Test, was organized to validate and compare these treatments.

\begin{figure}[t] 
\centering
\includegraphics[trim=0in 0in 0in 0in, clip=true, clip=true, width=0.5\textwidth]{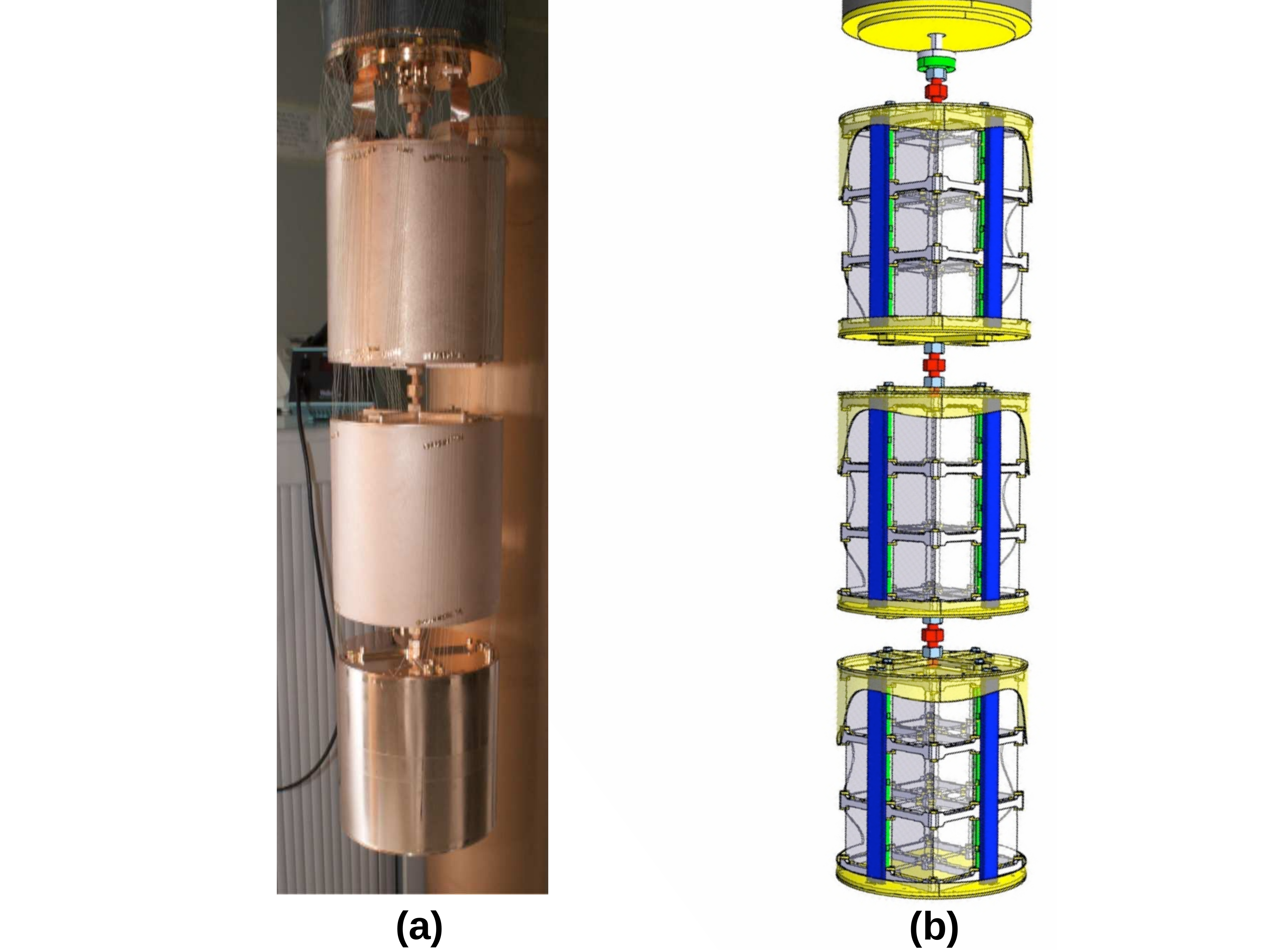}%
\caption[Three Towers Test detectors]{Photograph of the Three Towers Test detector with copper box (a) and picture of the detector without copper coverage (b).
From top to bottom: the polyethylene tower (T1), the LNGS tower (T2), and the Legnaro tower (T3).}
\label{fig:ttt_hanging_towers}
\end{figure}

\begin{figure}[t] 
\centering
\includegraphics[trim=0in 0in 0in 0in, clip=true, width=0.15\textwidth]{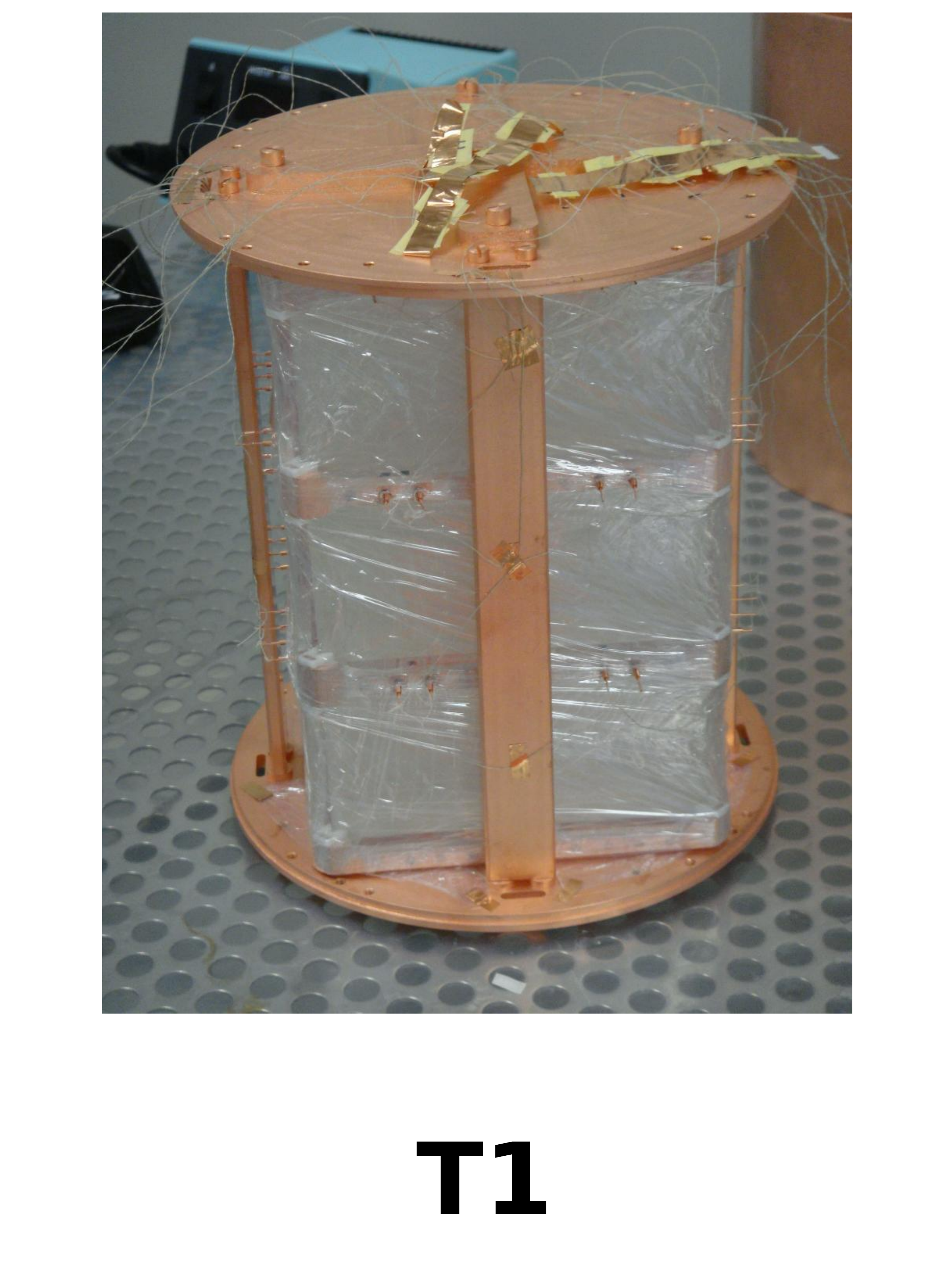}\label{fig:ttt_polyethylene_tower}%
\hspace{0.01\textwidth}%
\includegraphics[trim=0in 0in 0in 0in, clip=true, width=0.15\textwidth]{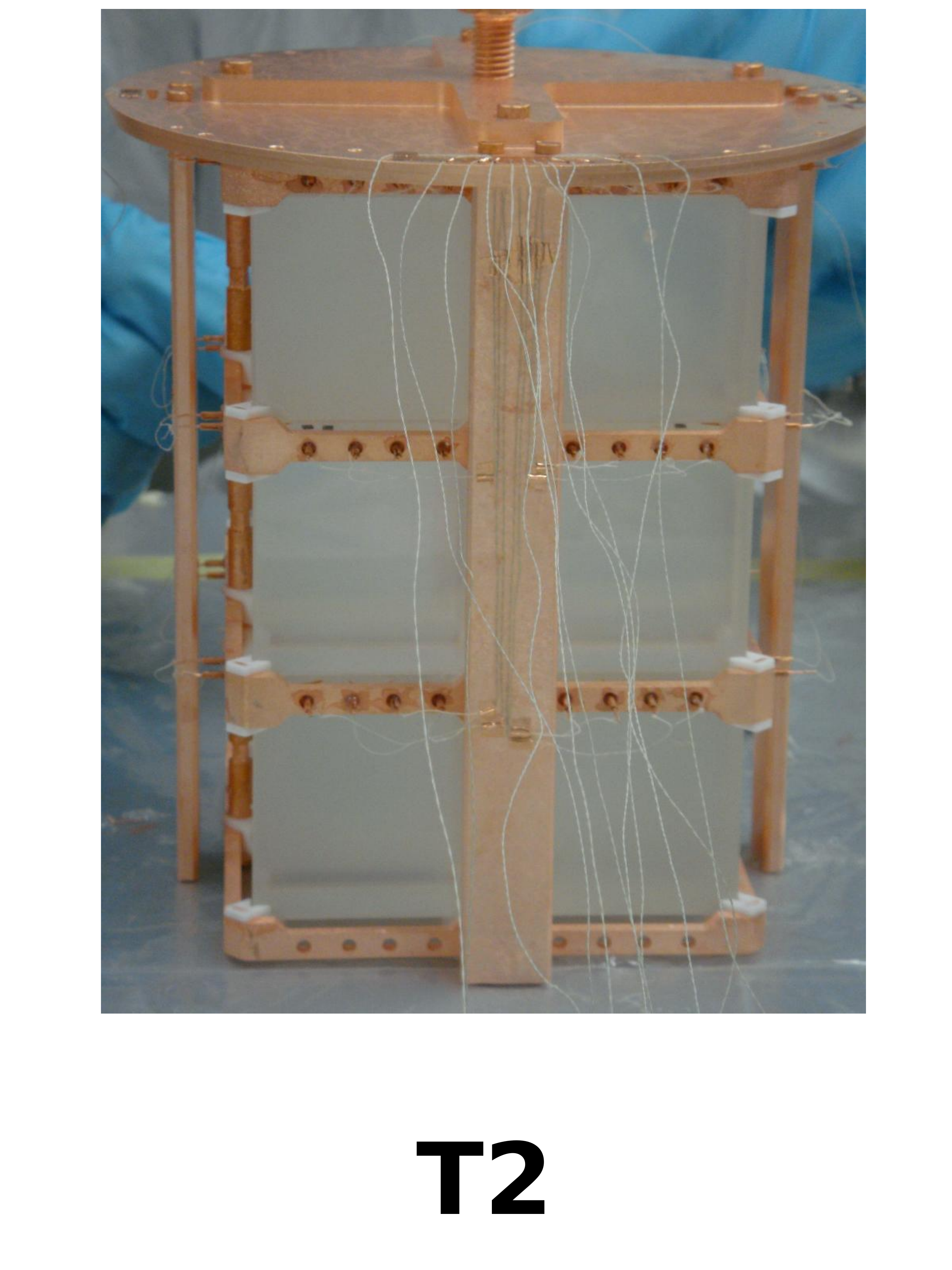}\label{fig:ttt_lngs_tower}%
\hspace{0.01\textwidth}%
\includegraphics[trim=0in 0in 0in 0in, clip=true, width=0.15\textwidth]{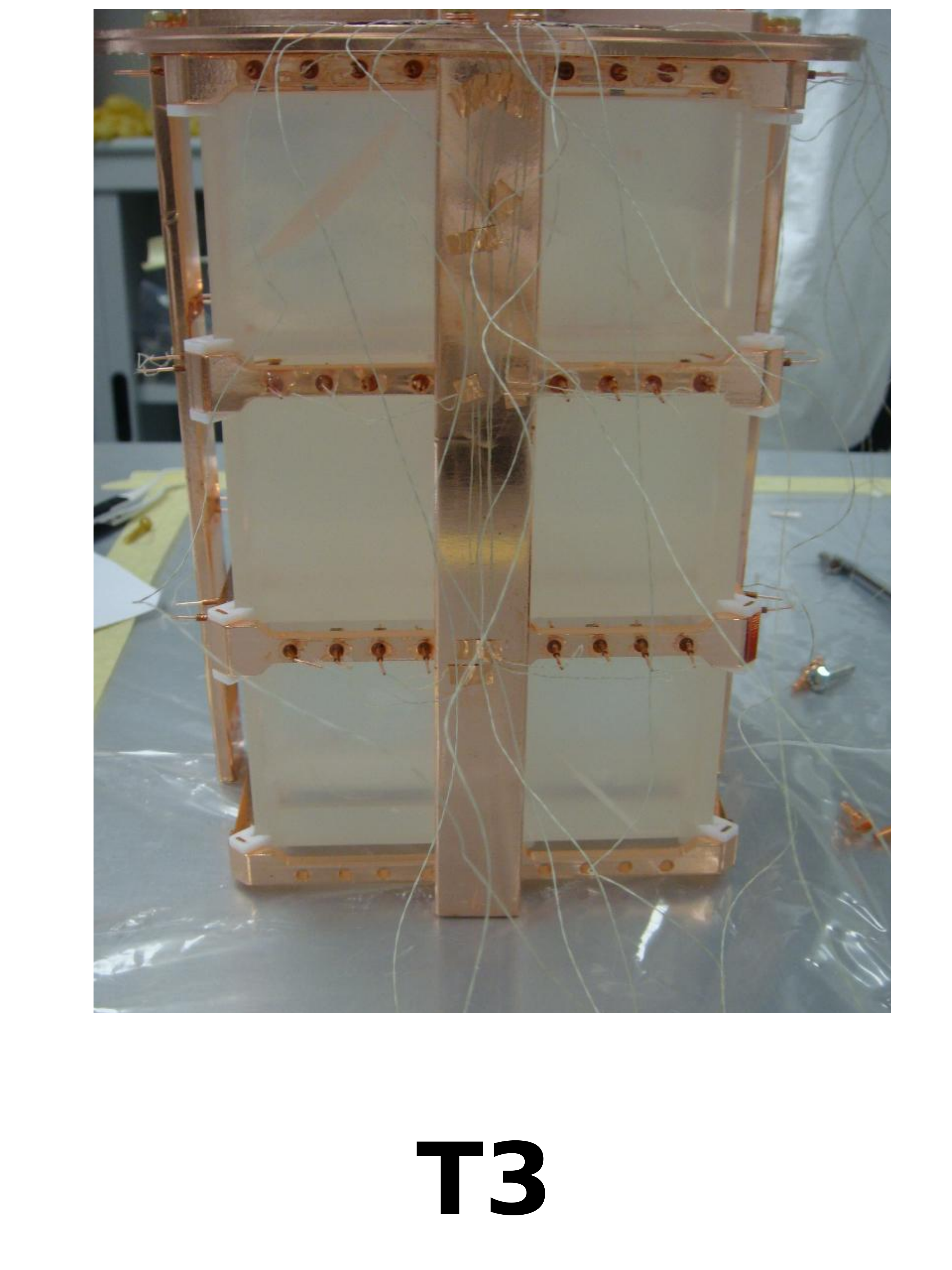}\label{fig:ttt_legnaro_tower}%
\caption[Three Towers Test detectors without cylindrical shields]{Photographs of the three 12-crystal arrays (the three towers ) without their cylindrical shields:  (a) the polyethylene tower (T1), (b) the LNGS tower (T2), and (c) the Legnaro tower (T3).}
\label{fig:ttt_detectors_open}
\end{figure}


\section{The Three Tower Test (TTT)}
\label{sec:exp}
%
The apparatus, known as the TTT detector, consisted of three 12-crystal arrays separated from each other by copper shields, see Figures~\ref{fig:ttt_hanging_towers} and \ref{fig:ttt_detectors_open}. The 36 crystals were all taken from Cuoricino production. Before being installed in the TTT detector their surfaces were retreated using new procedures developed for CUORE~\cite{11CCVR}. Great care was taken to ensure the crystals had similar contamination levels, history, and treatment. 
The copper of the three towers was taken from one single batch of NOSV copper produced by the Norddeutsche Affinerie, and machined following identical procedures. The surfaces of the copper parts  were treated with one of the three following procedures~\footnote[3]{The specifications of the reagents used in the threatments are: soap - Micro 90, Sigma Aldrich~\cite{Sigma};  phosphoric acid - 85\%, RPE analytical grade, Carlo Erba~\cite{CarloErba}; H$_2$O - obtained with Milli-Q system~\cite{MilliQ}, 18.2 M$\Omega$*cm at 25$^o$C; solid citric acid - 99\%, Sigma Aldrich~\cite{Sigma}; nitric acid - 67-69\% RS hyper-pure distilled with DuoPur purification system from Milestone, Carlo Erba~\cite{CarloErba}; Butanol - RPE analytical grade, Carlo Erba~\cite{CarloErba}.}:

\begin{itemize}
\item {\bf T1}: For the first or top tower, all the copper pieces were simply cleaned with soap, treated with a mixture of H$_2$O$_2$, H$_2$O and citric acid, and then wrapped with a few layers, \ca 70 $\mu$m thick on average, of polyethylene film. Given that the range of $\alpha$ particles in  polyethylene is about 20 $\mu$m, this is sufficiently thick to stop any $\alpha$ emission from the copper surface.  Commercial polyethylene film was chosen because it is relatively easy to handle; measurements of samples showed good radiopurity, $<$100 ppt of \udt and $<$700 ppt of \thdtn; and the procedure achieved good results in previous tests.  
\item {\bf T2}: For the second or middle tower, a purely chemical process with ultra-pure reagents was used. The first step was simple cleaning with soap and water. The copper pieces were then subjected to electroerosion with 85\% phosphoric acid, 5\% butanol, and 10\% water, followed by chemical etching with hyper-pure nitric acid. Finally a passivation step with H$_2$O$_2$, H$_2$O and citric acid was performed. The procedure was developed at LNGS on the basis of~\cite{Wojcik07}.
\item {\bf T3}: For the third or bottom tower, the LNL cleaning procedure was modified to include: Tumbling, Electropolishing, Chemical etching and Magnetron plasma etching (TECM). Final extraction and the following storage of parts was performed in a clean room to avoid re-contamination of the magnetron treated surfaces. Ultra clean reagents were not employed since the expense of supplying such a large TECM plant was deemed unsustainable and the procedure is designed for zero-deposition of foreign material on the treated piece.  
\end{itemize}

The TTT apparatus was installed in the cryostat used for Cuoricino~\cite{08arnaboldi}, the setup was identical to Cuoricino with respect to cryogenics, shields and front-end electronics, see references~\cite{08arnaboldi} and~\cite{10andreotti}. The data acquisition system (DAQ) designed for CUORE was used. The DAQ digitizes pulses from each bolometer front-end channel with an 18-bit ADC sampling at 125 Hz. The system acquires 626 samples, corresponding to 5.008~s, for each triggered pulse.  Data were collected in two campaigns: data-set 1 (Ds1) which ran from September  to  October 2009 and data-set 2 (Ds2) which ran from October 2009 to mid January 2010. During Ds1 a \kq source was present to validate parts of the offline analysis, this source was removed for Ds2.  The detector was operated at a higher energy threshold during Ds1 to reduce the trigger rate from the \kq source to an acceptable level.

Unfortunately not all the electrical connections survived the apparatus cool down, ultimately 12, 7, and 7 crystals were readable in T1, T2 and T3 respectively. The readable detectors performed quite well, the average FWHM resolution measured from the total spectrum (i.e. no multiplicity cuts applied) in each tower is shown in Table~\ref{tab:FWHM} for both data sets.  Three energy regions are considered: (i) low energy, i.e. the 352~keV $^{214}$Pb peak; (ii) an intermediate region near the ROI, i.e. the 2614.5~keV $^{208}$Tl peak; and (iii) the $\alpha$ region, i.e. the 5407~keV \podd peak. The first region was below threshold for Ds1.  The poorer average resolution which is evident at higher energy, the $\alpha$ region, is not due to the resolution of the individual bolometers but rather to the inter-bolometer calibration within a single tower. The highest available calibration peak is in fact 2614.5~keV, beyond this point the calibration must be extrapolated.

\begin{table}[t]
\begin{center}
\begin{tabular}{|c|c|c|c|c|}
\hline
{\bf Tower} & {\bf Ds} & \multicolumn{3}{c|}{\bf FWHM [keV]}\\
{}	& {}	  &{\bf @ 352~keV}	   &{\bf @ 2.6~MeV}      &{\bf @ 5.4~MeV} \\
\hline
\hline
{\bf T1}	& 1	  &	-		   &   3.0 \pom 0.5	    &	  23 \pom 5 \\
{\bf T2}	& 1	  &	-		   &   4.5 \pom 1.2	    &	  12 \pom 3 \\
{\bf T3}	& 1	  &	-		   &   5.2 \pom 1.0	    &	  11 \pom 3 \\

\hline
\hline
{\bf T1}	& 2  & 2.6  \pom  1.0   	     & 6.2   \pom  1.0      &  21 \pom 4\\
{\bf T2}	& 2	  & 1.6  \pom  0.3 	     & 3.7   \pom  0.7      &  27 \pom 5\\
{\bf T3}	& 2	  & 3.6  \pom  1.0 	     & 4.2   \pom  1.5      &  17 \pom 4\\
\hline
\end{tabular}
\end{center}
\caption[Average FWHM measured by the Three Towers]{Average FWHM measured by the Three Towers in the two data sets.}
\label{tab:FWHM}
\end{table}

\subsection{Data analysis}
\label{sec:ana}

Data from the TTT were analysed using the software framework developed for CUORE and already successfully used for Cuoricino~\cite{10andreotti}.  The off-line analysis identifies and rejects occasional periods with excess noise, applies pulse-shape analysis to identify and remove spurious pulses, and evaluates pulse height. Pulse height is a proxy for energy, the precise relationship between pulse height and energy is fixed using calibration sources.
The off-line analysis also identifies time coincidences between events for multiplicity analysis.  Ultimately the multiplicity analysis was only effective for T1 during Ds2 because this was the only tower where all 12 crystals were readable and the event rate from the \kq source was too high during Ds1.

The algorithms used for TTT analysis were identical to those described in~\cite{10andreotti} except a new method used for detector response stabilization~\cite{10Vignati}. Detector response stabilization is a correction applied to pulse amplitudes to account for gain instabilities and temperature fluctuations. The new algorithm  was validated using the 1461~keV $\gamma$-peak from the \kq source data in Ds1.


For each tower the average 
efficiencies are evaluated independently for the two data-sets following the procedure described in~\cite{10andreotti}. All the rates discussed in the following Sections are corrected for the efficiency. 

Table~\ref{tab:LT} summarizes the number of working detectors, the mass$\times$time exposure, the thresholds and the possibility of applying a multiplicity analysis for each data set.

Next we introduce four classes of energy spectra which are important for the data analysis.
\begin{itemize}
\item{\bf {\em Total} spectrum:} Each event is recorded at the corresponding energy in the spectrum regardless of any coincidence between crystals. 
\item{\bf M1 spectrum:} Events are selected if no other event occurred in the other bolometers in the tower within the coincidence window. A multiplicity=1 is assigned to such events. This is also referred to as the anti-coincidence spectrum.  
\item{\bf M2- and M2Sum-spectrum} An event is selected if there is exactly one more event in one of the other bolometers in the tower within the coincidence window. A multiplicity=2 is assigned to the two events and they are said to satisfy the coincidence cut. In the M2 spectrum the individual energies are recorded separately. In the M2Sum spectrum an entry is recorded at the energy corresponding to their sum.
\item{\bf M$>$2:} Multiple coincidence cut. An event is selected if there are two or more events in the other bolometers in the tower within the coincidence window.
\end{itemize}

\begin{table}[t]
  \begin{center}
    \begin{tabular}{|c|c|c|}
      \multicolumn{3}{c}{\bf Working detectors}\\
      \hline
      {\bf Tower}		&  {\bf Ds1}  &{\bf Ds2}\\
      \hline
      {\bf T1}		&    12     & 	   12 		      \\
      {\bf T2}		&    7     & 	   7 		      \\
      {\bf T3}		&    7     & 	   7 		      \\
      \hline
      \multicolumn{3}{c}{\bf Exposure [kg$\times$y]}\\
      \hline
      {\bf Tower}		&  {\bf  Ds1}  &{\bf Ds2}\\
      \hline
      {\bf T1}		&    0.87     & 	   1.09 		             \\
	{\bf T2}		&    0.51     & 	   0.76 		       \\
{\bf T3}		&    0.52     & 	   0.74 		       \\
\hline
\multicolumn{3}{c}{\bf Threshold [keV]}	\\
\hline
{\bf Tower}		&  {\bf  Ds1}  &{\bf Ds2}\\
\hline
{\bf all}                   	    & 500	&     100			\\
\hline
  \multicolumn{3}{c}{\bf Multiplicity}	\\
 \hline
 {\bf Tower}		&  { \bf Ds1}  &{\bf Ds2}\\
 \hline
 {\bf  all}	                           &  NO	&    only T1  		\\
 \hline
 \end{tabular}
 \end{center}
 \caption[Exposure]{Number of working detectors, exposure, energy  threshold and enabling of multiplicity analysis for the three towers  in the two data sets (Ds1 and Ds2).}
 \label{tab:LT}
 \end{table}   
 

\begin{table}[t]
\begin{center}
\begin{tabular}{|c|c|c|}
\hline
 {\bf Tower}          &{\bf Th}	                  &{\bf U}\\
                      &{ 238~keV peak}	          &{ 352~keV peak}\\
		      &[counts/h]$\times$10$^{-3}$	          &[counts/h]$\times$10$^{-3}$ \\
\hline
\hline
{\bf T1}	      &5.6   \pom   2.3   &   5.4   \pom   2.1\\
{\bf T2}	      &9.5   \pom   3.5   &   6.6   \pom   2.7\\
{\bf T3}	      &5.4   \pom   2.5   &   3.1   \pom   1.6\\
\hline
\end{tabular}
\end{center}
\caption[Intensity of low energy $\gamma$ lines]{Intensity of the 238~keV and 352~keV peaks (due to \thdt and \udt chains respectively) in the three towers. No coincidence cuts have been applied. Only Ds2 has been used.}
\label{Tab:LowGammas} 
\end{table}

Interpretation of the data in terms of higher level quantities such as the surface distribution of impurities relies heavily on Monte Carlo simulation.  We use the Geant4 toolkit~\cite{geant} with our detector geometry implemented to track particles, their interactions and energy deposition. Event generation relies on a package called GENDEC, developed and tested for previous \teod bolometric arrays~\cite{09bucci}, to generate particles with the correct time correlation along the decay chains of isotopes of interest. Another package called G2TAS~\cite{09bucci} simulates specific acquisition and analysis features of bolometers arrays such as pile-up, event multiplicity, and detector thresholds and resolutions. The complete simulation chain used in this work is identical to that described in detail in~\cite{09bucci}, except that we adopted the recently developed Livermore Physics Lists~\cite{11Livermore_PL}, which better implements low energy interactions. In~\cite{09bucci} it was shown that calibration measurements obtained by exposing a \teod bolometric array to a \thdt wire source were reproduced at better than 5\% with MC simulations. Given that the uncertainties in the Geant4 physics models are estimated to be about 5\%~\cite{pandolaMC} and particle transport, interaction, and detection add another few percent, on the whole we consider  10\%  as a conservative estimate for systematic uncertainty in the simulation output.

\begin{figure*}[t] 
\centering
\includegraphics[trim=0in 0in 0in 0in, clip=true, width=0.6\textwidth]{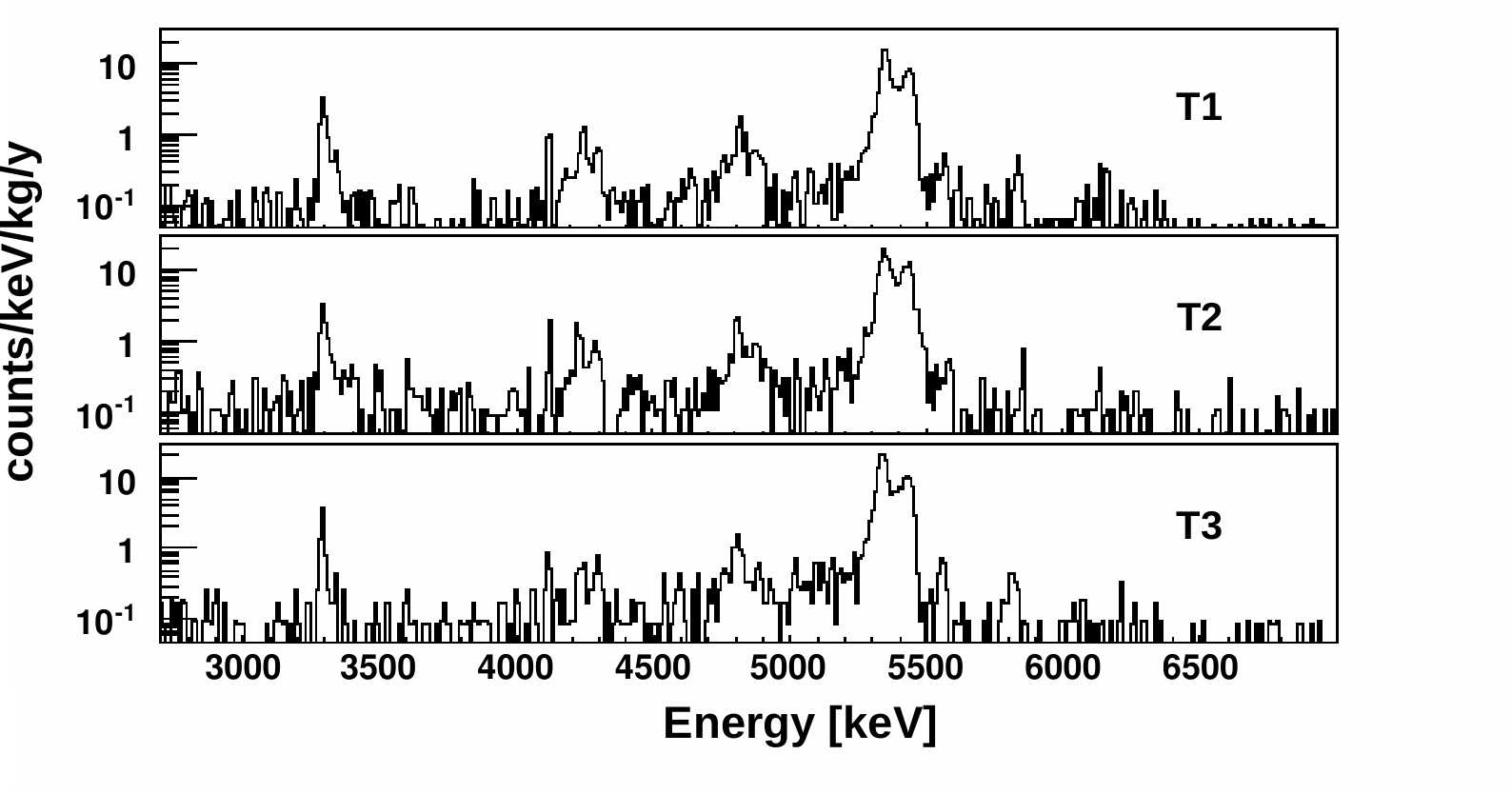}%
\caption[Comparison of the TTT measured spectra in the $\alpha$ region]{Comparison of the TTT spectra in the $\alpha$ region. T1 spectrum is shown in the top plot, T2 in the middle and T3 in the bottom one. No coincidence cuts are applied. The complete population of events has been used.}\label{fig:cfrTTT-T}
\end{figure*}

\begin{table*}[t]
\begin{center}
\begin{tabular}{|c|c|}
\hline
{\bf Tower}&{\bf continuum}        \\
	   &{2.7-3.9~MeV}  \\
	   &{[counts/keV/kg/y]}  \\
\hline
\hline	   
{\bf T1}	  &0.068	    \pom0.006 \\
{\bf T2}	  &0.120	    \pom0.012 \\
{\bf T3}   	  &0.072	    \pom0.008  \\
\hline
\end{tabular}
\end{center}
\caption[Counting rates in the 2.7-3.9~MeV energy region]{Counting rates [counts/keV/kg/y] recorded in the 2.7-3.9~MeV energy region with the three towers. Data refer to the sum of the two data sets, no coincidence cuts are applied.}
\label{tab:TTTRates} 
\end{table*}


\subsubsection{Comparison of surface radiation signatures measured in each tower}
\label{sec:resRates}

Three signatures are studied to constrain surface contamination with \udtn, \thdtn, and \poddn. Namely the degraded $\alpha$ signature between 2.7~MeV and 3.9~MeV described in Section~\ref{sec:MC}, the 238~keV $\gamma$-line from \thdt chain and the 352~keV $\gamma$-line from \udt chain. These low energy gammas have a short range (about 1 cm for a 300~keV gamma in copper) and thus events from these lines can only come from bulk and surface contamination of the crystals or from bulk and surface contamination of inert materials in their proximity. They do not have sufficient energy to reach the detectors from far inert materials.
    
Table~\ref{Tab:LowGammas}, shows the measured rates for the low energy $\gamma$-lines for Ds2; both lines were below threshold in Ds1. Based on HPGe measurements of copper and bolometric measurements of the crystals we estimate that less than 25\% comes from bulk contamination with \udt and \thdt (see Section ~\ref{sec:cuCUORE} and~\cite{11CCVR}). Therefore the majority of these events are due to surface radiation from the crystals or copper. The measured intensities in each of the towers are consistent within the uncertainty. 

Figure~\ref{fig:cfrTTT-T} shows the Total spectrum summed over Ds1 and
Ds2 for T1, T2 and T3. The rates in the degraded $\alpha$ window are reported in Table~\ref{tab:TTTRates}. The T1- and T3-rates are compatible within 1$\sigma$, whereas the rate in T2 is 70\% higher. To benchmark and compare the effectiveness of the copper surface treatments the relative contributions of crystal surface contamination vs. copper surface contamination to the observed Total rates must be quantified. To do this directly requires a multiplicity analysis; this is done for Ds2 of T1 in the next Section where we demonstrate that less than 20\% of the observed continuum rate is due to surface radiation from the crystals. Multiplicity analysis was not possible for T2 and T3 due to loss of channels during cooldown. 
Instead we rely on the fact that all 36 crystals, all belonging to the same production batch and treated identically, should have similar bulk and surface contamination levels. Thus while crystal surface contamination may contribute to the continuum counting rate, its variation between the towers should be dominated by differences in the residual surface radiation from the treated copper surfaces.

\begin{figure*}[tp] 
\centering
\includegraphics[trim=0in 0in 0in 0in, clip=true, width=0.75\textwidth]{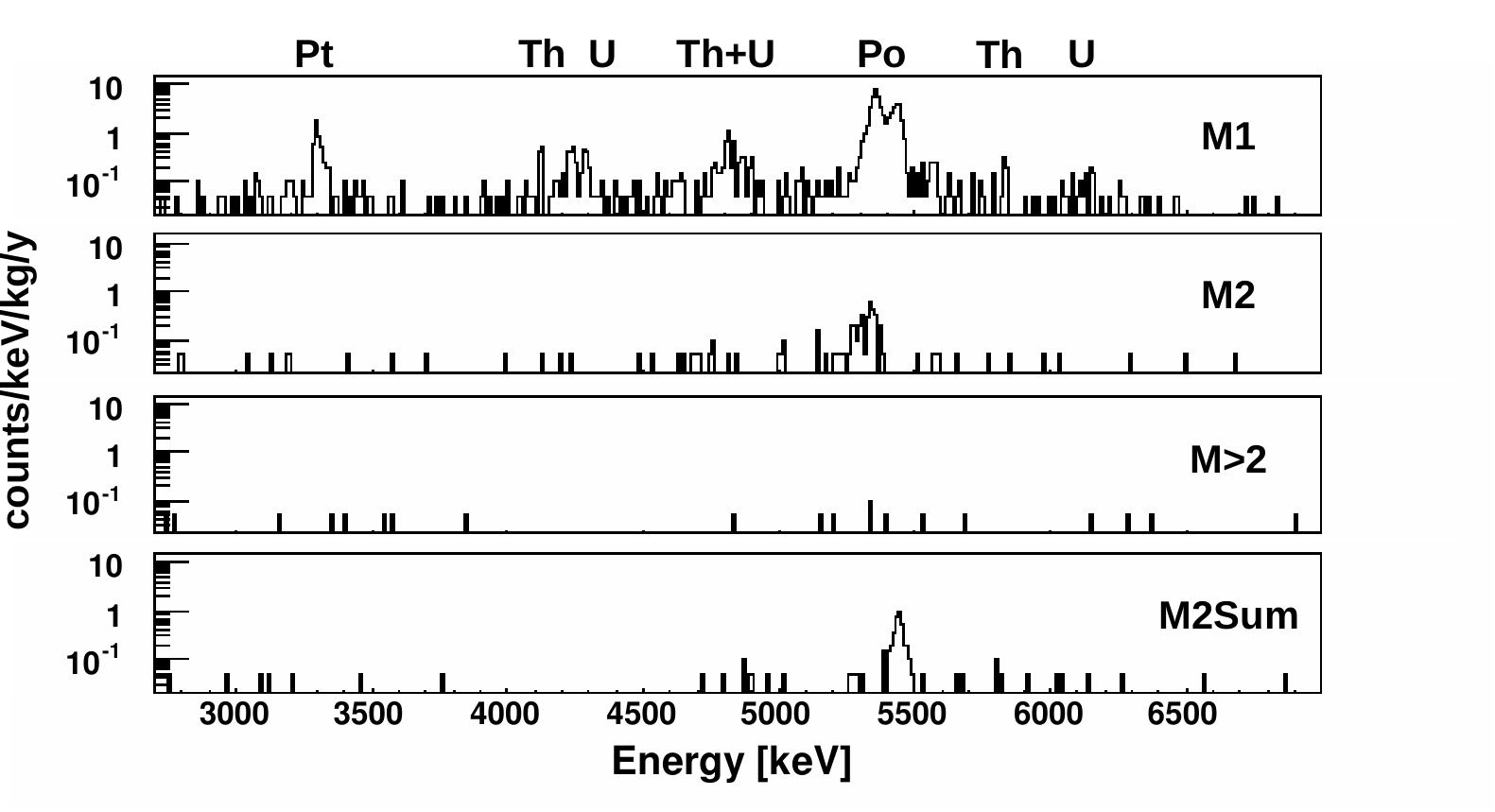}%
\caption[T1 multiplicity spectra in the $\alpha$ region]{M1 (at the top), M2 (the second from the top), M$>$2 (the third from the top) and M2Sum (the bottom one) energy spectra in the $\alpha$ region for T1 acquired events. The name of the parent of the decay chain is written above highest peaks.}
\label{fig:T1MSpectra}
\end{figure*}

\begin{table*}[t]
\begin{center}
\begin{tabular}{|c|c|c|c|c|}
\hline
{\bf Mult.}&{\bf continuum}           &{\bf U/Th}              &{\bf Po}	     &{\bf U/Th}   \\   
	   &{2.7-3.9~MeV}             &{4-5~MeV}               &{5-6~MeV}	     &{6-8~MeV}     \\ 
	   &{[counts/keV/kg/y]}       &{[counts/keV/kg/y]}     &{[counts/keV/kg/y]}  &{[counts/keV/kg/y]}\\ 
\hline	   
{\bf Total}    	&0.068		\pom0.009		&0.29	   \pom0.02	&1.38	  \pom0.07    &    0.03    \pom 0.005\\
{\bf M1}   	&0.052		\pom0.008		&0.28	   \pom0.02	&1.30	  \pom0.07    &    0.02    \pom 0.004\\
{\bf M2}   	&0.009		\pom0.003		&0.0018	   \pom0.005	&0.09	  \pom0.01    &    0.005   \pom 0.002\\
{\bf M$>$2}     &0.008		\pom0.003		&0.001     \pom0.001	&0.008	  \pom0.003   &    0.005   \pom 0.002\\

\hline
\end{tabular}
\end{center}
\caption[T1 counting rates in the $\alpha$ region with different multiplicity constraints]{Rates [counts/keV/kg/y] measured in T1 in the $\alpha$ region with different multiplicity constraints. Only the statistics collected in the second part of the measurement is used.}
\label{tab:T1MultRates} 
\end{table*}

\subsubsection{Constraints on the contribution of crystal surface radiation to the 2.7-3.9~MeV rate for T1}
\label{sec:T1}

The M1, M2, M2Sum and M$>$2 spectra used for the multiplicity analysis are shown in Figure~\ref{fig:T1MSpectra}.  

In the M1 spectrum we note the following features: the \ptc peak at 3270~keV, peaks between 4 and 5~MeV identified with $\alpha$ transistions in the chains of \udt and \thdtn, and peaks at about 5407~keV from $\alpha$ decay of \podd.  The only peak centered at the energy of the $\alpha$ fragment of any of the $\alpha$ transitions considered is the peak at 5304~keV.  This peak, whose count rate is almost constant in time, is consistent with a very thin layer (of the order of 0.01 $\mu$m) of \podd contamination on the surface of the crystals or inert materials facing them such that the $\alpha$ is detected in a facing bolometer without any loss of energy while the recoiling nucleus remains in the source element.  The clear rate discrepancy between the intensity of the \podd $\alpha$ peaks and those identified with the \udt chain indicates that the \podd contamination is not primarily supported by \udtn. A break in the secular equilibrium in the lower part of both the \udt and \thdt chains is also evident from the discrepancy between the intensity of the peaks above 6~MeV and those in the 4-5~MeV range (see also rates in Table ~\ref{tab:T1MultRates}).

In the M2 spectrum the only visible structure is at 5304~keV. It is an asymmetric peak, with a tail on the left side.  This is consistent with \podd decays occurring in a shallow layer on the crystal surface where the $\alpha$ particle and recoiling nucleus release energy in two facing detectors. 

In the M2Sum spectrum the most clear peak is at 5407~keV. This peak is a different signature of the same \podd contamination on the crystal surface responsible for the 5304~keV line in the M2 spectrum.
Using a MC simulation we extrapolate the intensity of the M2Sum peak
at 5407~keV to estimate the contribution of crystal surface contamination to the M1 peak at 5304~keV.  The extrapolated rate is about a factor 5 lower than the measured one.  Thus  most of the events in the 5304~keV peak of the M1 spectrum are from \podd contamination at a shallow depth in the inert material facing the crystals, mainly polyethylene in the case of T1. 

In the M$>$2 spectrum no peaks are evident. Only a few sources are expected to produce high-multiplicity multiplets of events with a continuum energy spectrum. For example muon cascades, neutron interactions, $\gamma$, and $\beta$+$\gamma$ from decays with high Q-values (like \tld $\beta$ decay whose Q-value is of about 5~MeV) taking place in the detector proximity.

Table~\ref{tab:T1MultRates} lists the rates measured in the degraded-$\alpha$ region in Ds2 after the different multiplicity cuts. In addition, the results for three higher energy intervals are shown. These intervals cover transition energies of $\alpha$ decays in the chains of \mbox{\udtn, \thdt (4-5~MeV and 6-8~MeV)} and \podd (5-6~MeV) and their daughters.  Thus events in these intervals give a first indication of the presence of a contamination in the crystals bulk and/or surface.
As explained in Section~\ref{sec:MC}, a contribution to the peaks at the $\alpha$ energy in the M1 spectrum could arise also from surface contamination distributed in a very shallow depth of the facing materials, but this is expected to give a minor contribution. 

Following the methodology in~\cite{11CCVR}, we use simulation to estimate the maximum contribution to the 2.7-3.9~MeV region  from \udtn, \thdt and \podd impurities on the crystal surfaces.  For each chain, the simulation is used to predict the Total, M1, M2 and M2Sum spectra for an assumed surface distribution of the impurity.   The surface impurity distribution is taken of the following form:  $\rho = \rho_0  e^{-x/\lambda}$,  where $\rho_{0}$ is the contamination per unit volume at the surface, $x$ is the distance into the crystal, and $\lambda$ is the mean penetration depth of the impurity.  For a given $\lambda$ the comparison between MC simulation and measurement rates fixes the normalizations of the Total, M1, M2 and M2Sum simulated spectra for each source. The integral of $\rho$ over the volume, multiplied for the evaluated normalization factor and divided by the surface of the source element (on which the distribution is assumed to be uniform) gives what we quote as surface contamination $\delta$, i.e. the concentration of impurities measured in Bq/cm$^2$.

For the \udt and \thdt chain, with $\lambda$ fixed, we estimate the maximum $\delta$ consistent with the rates measured in the M1- and M2Sum-spectra at the Q-values of the $\alpha$ transitions in the chain.  Where no peaks were visible in the spectrum, the 90\%~C.L. Feldman-Cousins upper limit~\cite{FC} is evaluated using the number of counts recorded in a 60~keV energy interval centered at the transition Q-value and assuming a background of zero counts~\footnote[4]{The choice of considering 0 background counts leads to an overestimation of the crystal contamination and therefore to a more conservative evaluation.}. We repeat this procedure for a range of values of $\lambda$ between 0.01~$\mu$m and 5~$\mu$m. A similar procedure is used for \poddn, except that the estimate for $\delta$ is only required to be consistent with the rate measured at the 5.4~MeV peak in the M2Sum spectrum.   
Depending on the choice of $\lambda$, we find values for $\delta$ ranging from 3$\times$10$^{-9}$~Bq/cm$^2$ to 9$\times$10$^{-9}$~Bq/cm$^2$ for \udt and \thdtn, and from 4 $\times$10$^{-8}$~Bq/cm$^2$ to 8$\times$10$^{-8}$~Bq/cm$^2$ for \poddn . 

Given a pair ($\lambda$, $\delta$), the expected count rate in the degraded $\alpha$ window in the M1 or Total spectrum is fixed by the  simulation. The maximum expected count rate from \udtn, \thdt and \podd combined ranges between 2\% and 10\% of the measured rate for the M1 spectrum and between 3\% and 20\% of the measured rate for the Total spectrum.  Again the range corresponds to the interval explored for $\lambda$, with the lowest percentage corresponding to the smallest $\lambda$. We conclude that less than 20\% of the rate measured between 2.7-3.9~MeV in the Total spectra for the towers (see Table~\ref{tab:TTTRates}) is from crystal surface contamination. 

For two reasons we neglect the possible contribution of environmental neutrons and muons to the difference in measured rates.  Firstly neutron and muon backgrounds should be the same for all three towers since they are in the same local environment. Secondly, on the basis of data from previous \teod experiments the expected contributions to the degraded $\alpha$ window of the Total spectrum are negligibly low:  the 90\%~C.L. upper limit for the expected neutron contribution is 8$\times$10$^{-6}$~counts/keV/kg/y~\cite{09bucci} and the expected contribution from muons is (1.26 $\pm$ 0.3)$\times$10$^{-3}$~counts/keV/kg/y~\cite{10andreotti_mu}. 
The remaining candidate source is surface contamination of inert materials surrounding the crystals, mainly the copper holder for T2 and T3 and the multiple layers of polyethylene film for T1.

Since the rates measured in T1 and T3 are comparable we conclude that the polyethelyene and TECM treatments are similarly effective. The treatment for T2, which had a 70\% higher rate than the other two is the least effective.


\subsubsection{Copper surface contamination derived from the measured Total rates}
\label{sec:CuLimits}

In this Section we proceed with the evaluation of the surface contamination of the copper facing the detectors. We do this only for T2 (chemical etching) and T3 (TECM) where the copper faces the crystals without any interposed material.

We focus on the following signatures to estimate the concentration of $^{238}$U, \thdt and \podd impurities in the surface of the copper holder from the measured spectra: (i) counting rate in the degraded $\alpha$ window, (ii) the 238~keV and 352~keV peak intensities, (iii) the intensities of $\alpha$ peaks centered at the $\alpha$ energy (transition energy - nuclear recoil energy). None of these signatures is exclusively due to contamination of the copper surface (for example impurities of the crystals themselves could contribute) and in particular, in the case of the degraded $\alpha$ window, more than one species can contribute. For a quantitative evaluation we assume that the 100\% of the measured rate is due to each species of impurity on the copper surface, $^{238}$U, \thdt and \poddn, in turn.  The surface concentrations thus derived are conservative upper limits. The main steps of the analysis are the following:  
\begin{itemize}
\item For each species we simulate events assuming an exponential depth profile for the contamination as described in the previous Section. We consider different characteristic penetration depths, $\lambda$, ranging from 0.1 $\mu$m to 10 $\mu$m. The former is the most shallow distribution whose effects can be experimentally identified and the latter is the range of a 5~MeV $\alpha$ in copper. Impurities at greater depths are indistinguishable from bulk contamination.
\item For each of the considered signatures we evaluate the detection efficiency, i.e. the fraction of simulated decays that result in an event with the signature of interest. We then identify the most efficient signature of the contaminant being studied.
\item For the most efficient signature the maximum activity compatible at 90\%~C.L with the experimentally measured counting rate is calculated.  The copper activity per unit area is then calculated by dividing this result by the emitting surface area.
\end{itemize}
Since the actual characteristic depth is unknown, we report in Table~\ref{tab:CuCont} the contamination upper limits corresponding to the depth profile which gave the worst, i.e. the highest, value. The meaning of this upper limit is that in the worst configuration the copper surface contamination is below the reported activity at the 90\%~C.L.

\begin{table}[t]
\begin{center}
\begin{tabular}{|c|c|c|}
\hline
{\bf Source}&{\bf T2 [Bq/cm$^2$]}&{\bf T3 [Bq/cm$^2$]}\\
\hline
\hline  	  
{\bf \thdt}&			 1.3$\times$10$^{-7}$ &    6.8$\times$10$^{-8}$\\
\hline
{\bf \udt}&			 1.3$\times$10$^{-7}$ &    6.5$\times$10$^{-8}$\\
\hline
{\bf \podd}&			 8.8$\times$10$^{-7}$ &    8.6$\times$10$^{-7}$\\
\hline
\end{tabular}
\end{center}
\caption[Copper surface contamination limits]{Evaluated 90\%~C.L. upper limits for T2 and T3 copper surface contamination.}
\label{tab:CuCont} 
\end{table}

\section{Contribution of copper to the ROI for CUORE}
\label{sec:CUORE}

An important final step is to estimate the residual background rate in the ROI for CUORE to confirm if it is compatible with the goal of the experiment (i.e. $\leq$10$^{-2}$~counts/keV/kg/y). Assuming a completely functioning array, which will allow full exploitation of multiplicity analysis, the figure of interest for CUORE is the background in the  ROI of the anti-coincidence or M1-spectrum.

For a given tower of TTT the main observable used to benchmark the treatment is the rate in the degraded $\alpha$ window (2.7 to 3.9~MeV) of the Total spectrum. 
To estimate the transfer coefficient that connects this observable to the the M1-rate of the CUORE spectrum in the ROI we turn again to MC simulation. For a given distribution of impurities on the surface of the copper structure we simulate datasets for both the TTT tower and CUORE and extract both rates of interest, i.e. the rate in the 2.7-3.9~MeV window of the Total spectrum for TTT and the rate in the \BBz-ROI of the M1 spectrum for CUORE. The ratio of these rates is an estimator for the transfer coefficient for the associated surface distribution.  Due to the completness of the simulation this estimator simultaneously includes: the geometrical effects, i.e the fact that on average less copper faces the crystals in CUORE than in TTT; the anti-coincidence effects,  i.e the fact that CUORE will benefit from full exploitation of multiplicity analysis; and spectral effects, i.e the fact that the degraded alpha spectrum increases approximately linearly with energy and thus has a lower contribution in the \BBz-ROI relative to the 2.7-3.9~MeV window.  
Given  the transfer coefficient the associated ROI-background in CUORE can be estimated from the experimentally measured rate in the degraded alpha window of the TTT tower.  
Transfer coefficients were estimated separately for \udtn, \thdt and \podd surface distributions and in each case, as before,  a range of values for the mean penetration depth $\lambda$ are used.  From this set of transfer coefficients we choose the one that gives the worst ROI-background for CUORE .

Following this procedure we estimate the ROI background in CUORE for the polyethelyene treatment, T1, and the TECM treatment, T3.  For T1, we subtract the crystal contribution estimated in Section~\ref{sec:T1} from the rate given in Table~\ref{tab:TTTRates}. For T3, we have no direct estimate of the crystal contribution so we assume conservatively that the measured rate is entirely due to copper surface impurities. The corresponding 90\%~C.L. upper limit for the background rate in the ROI of CUORE after the anti-coincidence cut is 0.02~counts/keV/kg/y for the T1 treatment and 0.03~counts/keV/kg/y for T3. If the crystal contribution in T3 could be considered to be the same of T1, then the 90\%~C.L. upper limit for the T3 treatment would also be 0.02~counts/keV/kg/y. These values represent for CUORE the major component of the background in the \BBz-ROI. All the other background sources are in fact expected to give contributions lower by about one order of magnitude (see Ref.~\cite{04arnaboldi}). 

\section{The first step towards CUORE: CUORE-0}

The first step towards CUORE is represented by CUORE-0. It consists of one CUORE-like tower, made of 13 planes of 4 \teodn, \ciccio crystals each, for a total mass of \ca11~kg of \tectn. CUORE-0 has been mounted inside the former Cuoricino dilution refrigerator and is intended to provide a test of proof for CUORE at all stages, while being a sensitive experiment itself, able to surpass Cuoricino in the \BBzn physics reach.
The crystals of CUORE-0 comes from the same production of CUORE; all detector components have been manufactured, cleaned and stored with the same protocols defined for CUORE. 

Based on the present results of the TTT array, and due to the greater simplicity compared to the polyethylene wrapping technique, the TECM procedure was chosen for the treatment of the surface of the copper skeleton. CUORE-0 is expected to take data for at least 2 years and will allow therefore to perform with a higher statistics an accurate analysis of the components constituting the ROI-background before CUORE starts the data-taking.


\section*{Conclusions}
The Three Towers Test was undertaken to validate and compare three different methods considered to control background due to surface radiation from copper parts in the CUORE detector. Looking for the best compromise between cost, reproducibility and background control, a mixed approach will be pursued for CUORE: small copper parts will be subjected to TECM while the large copper shields facing the external detector towers will be covered with polyethylene.  These two techniques gave compatible results for the event rate in the degraded $\alpha$ window, 2.7-3.9~MeV, without any coincidence cuts. 
The data from the TTT apparatus demonstrate that surface contamination levels lower than 7$\times$10$^{-8}$~Bq/cm$^2$ for \thdt and \udt and below 9$\times$10$^{-7}$Bq/cm$^2$ for \podd can be achieved.  This is the best purity level we ever achieved for copper surfaces. 
An extrapolation of the rate measured in the degraded $\alpha$ window of T1 and T3  to the ROI of CUORE was performed. Depending on a reasonable range of unknown parameters in the extrapolation model the 90\%~C.L. upper limit for the contribution to the ROI-background in the anti-coincidence spectrum of CUORE is between \ca0.02 and 0.03 ~counts/keV/kg/y. Since the surface contamination of inert materials facing the detectors constitutes, based on our current knowledge, the dominant contribution to the \BBzn background, this also represents the 90\%~C.L. upper limit for the ROI-background expected for CUORE. It translates into an upper limit at 68\%~C.L. of \ca0.01 ~counts/keV/kg/y, close to the design goal of the experiment and should allow CUORE to significantly improve limits on --- or perhaps discover --- Majorana neutrino masses.

\end{document}